\documentclass[conference]{IEEEtran}
\IEEEoverridecommandlockouts    
\usepackage{cite}
\usepackage{amsmath,amssymb,amsfonts}
\usepackage{algorithm} 
\usepackage{graphicx}
\usepackage{subcaption}
\usepackage{textcomp}
\usepackage{xcolor}
\usepackage{float}
\usepackage{caption}
\usepackage{amsmath}
\usepackage{algpseudocode} 
\usepackage{xcolor}

\newcommand{\ket}[1]{\left|#1\right\rangle}

\def\BibTeX{{\rm B\kern-.05em{\sc i\kern-.025em b}\kern-.08em
    T\kern-.1667em\lower.7ex\hbox{E}\kern-.125emX}}

\begin{document}

\title{Efficient Circuit Wire Cutting Based on Commuting Groups}

\author{\IEEEauthorblockN{Xinpeng Li}
\IEEEauthorblockA{\textit{Case Western Reserve University}\\
Cleveland, OH, USA \\
xxl1337@case.edu}
\and
\IEEEauthorblockN{Vinooth Kulkarni}
\IEEEauthorblockA{\textit{Case Western Reserve University}\\
Cleveland, OH, USA \\
vxk285@case.edu}
\and
\IEEEauthorblockN{Daniel T. Chen}
\IEEEauthorblockA{\textit{Brown University}\\ Providence, RI, USA \\
daniel\_t\_chen@brown.edu}
\and
\IEEEauthorblockN{Qiang Guan}
\IEEEauthorblockA{\textit{Kent State University}\\
Kent, OH, USA \\
qguan@kent.edu}
\and
\IEEEauthorblockN{Weiwen Jiang}
\IEEEauthorblockA{\textit{George Mason University}\\
Fairfax, VA, USA \\
wjiang8@gmu.edu }
\and
\IEEEauthorblockN{Ning Xie}
\IEEEauthorblockA{\textit{Florida International University}\\
Miami, FL, USA \\
nxie@cis.fiu.edu }
\and
\IEEEauthorblockN{Shuai Xu}
\IEEEauthorblockA{\textit{Case Western Reserve University}\\
Cleveland, OH, USA \\
sxx214@case.edu}
\and
\IEEEauthorblockN{Vipin Chaudhary}
\IEEEauthorblockA{\textit{Case Western Reserve University}\\
Cleveland, OH, USA \\
vxc204@case.edu}
\thanks{This research was supported in part by NSF Awards 2216923, 2117439, 2238734, 2217021 and 2311950.}
}

\maketitle
\begin{abstract}

Current quantum devices face challenges when dealing with large circuits due to error rates as circuit size and the number of qubits increase. The circuit wire-cutting technique addresses this issue by breaking down a large circuit into smaller, more manageable subcircuits. However, the exponential increase in the number of subcircuits and the complexity of reconstruction as more cuts are made poses a great practical challenge. Inspired by ancilla-assisted quantum process tomography and the MUBs-based grouping technique for simultaneous measurement, we propose a new approach that can reduce subcircuit running overhead. The approach first uses ancillary qubits to transform all quantum input initializations into quantum output measurements. These output measurements are then organized into commuting groups for the purpose of simultaneous measurement, based on MUBs-based grouping. This approach significantly reduces the number of necessary subcircuits as well as the total number of shots. Lastly, we provide numerical experiments to demonstrate the complexity reduction.

\end{abstract}

\section{Introduction}

Recent advancements in quantum computing and quantum hardware have had a transformative impact on various sectors in industry and academia~\cite{bayerstadler2021industry,bova2021commercial}. Despite the potential of quantum computing, many-qubit technologies still face great hurdles such as the limited number of qubits and the low fidelity of the qubits. IBM's creation of \textit{Condor}, one of the world's largest quantum computer systems with 1121 qubits~\cite{gambetta2020ibm}, highlights the great advancements of quantum hardware during the past decade. However, these systems are still far from practical applications.
John Preskill described such systems as Noisy Intermediate-Scale Quantum (NISQ) computers, emphasizing the significant work being done to improve qubit fidelity and increase qubit count \cite{LaRose2022mitiqsoftware, Abobeih_2022,saki2023hypothesis,gambetta_2022}. 

Given the constraints on the size of quantum hardware, how to effectively leverage the limited available quantum resources becomes a crucial problem. One promising approach is the quantum circuit cutting method, which divides a large quantum circuit into smaller circuit fragments that can be run independently on available quantum devices and the outputs of each fragments are combined classically. Empirical studies have shown that cutting a circuit helps reduce the impact of noise~\cite{ayral2020quantum, ayral2021quantum} and can be used for error mitigation~\cite{liu2022classical}. Additionally, research has focused on accurately accounting for statistical shot noise and adapting the approach to address specific challenges, such as combinatorial optimization~\cite{saleem2021quantum}. 
Therefore, this technique offers considerable potential for overcoming numerous practical challenges in utilizing quantum hardware, especially in the NISQ era.

However, the success of circuit cutting techniques are significantly hampered by the fact that the runtime may increase exponentially with the number of cuts. Similar to quantum state tomography, circuit cutting involves classically monitoring \emph{all} quantum degrees of freedom at the cut points, naturally leading to exponential blowup as the size of the quantum state grows. Efforts to reduce these costs include the implementation of randomized measurements~\cite{Lowe_2022, chen2022quantum}, classical sampling~\cite{Chen_2022}, golden circuit cutting points~\cite{chen2023efficient, chen2023online}, and variational optimization~\cite{Uchehara2022}. Nonetheless, the exponential growth in runtime is unlikely to vanish without imposing structural assumptions on the circuit. Alternatively, finding applications of circuit cutting that avoid the run time blowup issues, as demonstrated in~\cite{liu2022classical}, remains a question that is worth investigating.

The work introduced in this paper addresses the former issue of the exponential complexity of runtime.
Our method is built mainly upon the following two simple observations:

\textbf{Observation 1 (Initialization via Measurement Operator):} Inspired by ancilla-assisted quantum process tomography~\cite{altepeter2003ancilla, de2003exploiting}, we replace the states initialization at the front of the circuits by an equivalent procedure, which first entangles each input qubit with an ancilla qubit into a Bell state and then measures ancilla qubit on corresponding desired basis;

\textbf{Observation 2 (Simultaneous Measurement in a Commuting Group):} We utilize the well-known fact that observables within a commuting group—often referred to in the literature as a commuting family—can be diagonalized simultaneously. Therefore, measuring on any one operator provides the complete information of all the observables in the group.

Based on the observations above, we propose to replace quantum state initializations with measurements (Observation 1) by introducing additional ancilla qubits. 
Then, we employ simultaneous measurement (Observation 2) to efficiently measure all observables.
Recall that the Pauli operators (or \emph{Pauli strings}, which are the tensor products of $n$ Pauli matrices) provide a complete basis of observables on the Hilbert space of $n$ qubits. Thus, to obtain complete information for $n$ qubits, we must obtain measurement results from all $4^n$ Pauli operators. Partitioning these $4^n$ Pauli operators into groups can reduce the required measurements, thanks to simultaneous measurement. Therefore, a good partition is key to maximizing the benefits of simultaneous measurement.

Built on previous works in the framework of \emph{Mutually Unbiased Bases} (MUBs)~\cite{wootters1986quantum,schwinger1960unitary,ivonovic1981geometrical,wootters1989optimal, lawrence2002mutually}, Lawrence et al.~\cite{lawrence2002mutually} show that the optimal partition of $4^{n}-1$ Pauli operators is organizing them into $2^{n}+1$ commuting groups, each consisting of $2^{n}-1$ internally commuting observables. The existence of such optimal commuting operators partition implies, in the ideal case, polynomial improvements on the numbers of subcircuits a circuit cutting generates as well as the total number of shots required when measurement accuracy is taken into consideration. 
For realistic implementation, there are trade-offs: adding more ancilla qubits and introducing additional gates to the subcircuit.
Our experimental results demonstrate strong evidence of polynomial savings in the number of shots.

The rest of the paper is organized as follows. 
Background on circuit cutting, MUBs-based grouping and simultaneous measurement are reviewed in Section~\ref{se:Background}. 
We then discuss the methodology for converting initializations to measurements in Section~\ref{Ancilla assist}. Subsequently, we demonstrate our proposed method---combining techniques introduced in the previous sections---in Section~\ref{se:Circuit Wire Cutting based on commuting grouping} and we analyze the advantages and trade-offs of our method in the subsequent Section~\ref{se:pro con}. Finally, we validate our theory and analysis with empirical evidence in Section~\ref{se:experiment} and make conclusions in Section~\ref{conclusion}.


\section{Background}
\label{se:Background}
\subsection{Circuit Cutting}
\label{se:Circuit Cutting}
Circuit cutting is primarily classified into two methods: 'wire cutting' and 'gate cutting'~\cite{mitarai2021constructing,ufrecht2023cutting}. 
Circuit wire cutting can be further categorized into two types: with classical communication \cite{harada2303doubly,brenner2023optimal,lowe2023fast} and without classical communication \cite{peng2020simulating,tang2021cutqc}. 
In this paper, we focus on circuit cutting without classical communication. Hence, when we refer to 'circuit cutting', we are specifically discussing wire cutting without classical communication.


\begin{figure}[htbp]
    \centering
    \includegraphics[width=0.7\linewidth]{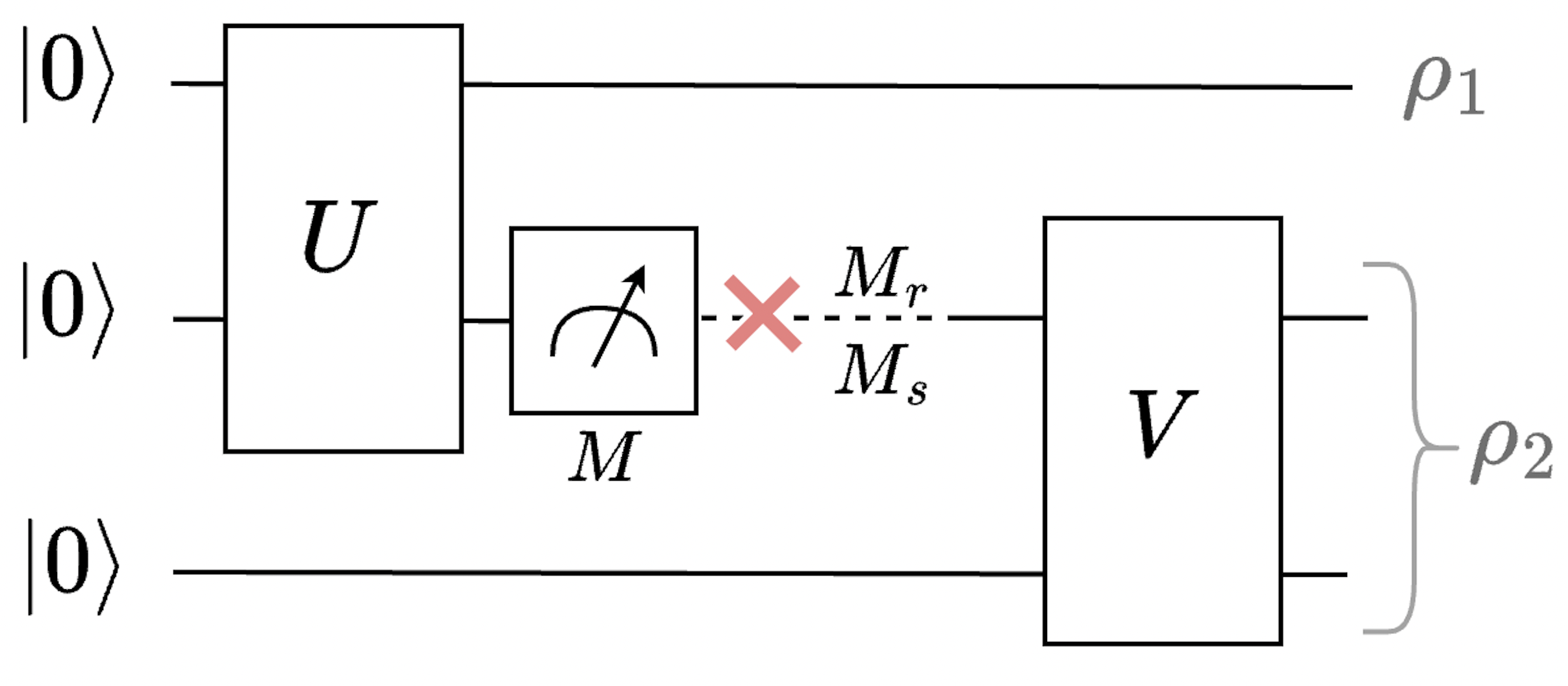} 
    \caption{An example of 3-qubit circuit cutting is depicted in \ref{eq:three qubit example circuit cutting}. The circuit is separated into two fragments, unitary $U$ and unitary $V$ by cutting at the a red cross mark. To construct the subcircuits, we need to measure the second qubit on the $M$ basis after $U$ and prepare the $M_r$ and $M_s$ states as the initial states for $V$. }
    \label{fig:circuit_cut_backgroud} 
\end{figure}

\begin{figure}[htbp]
    \centering
    \includegraphics[width=0.9\linewidth]{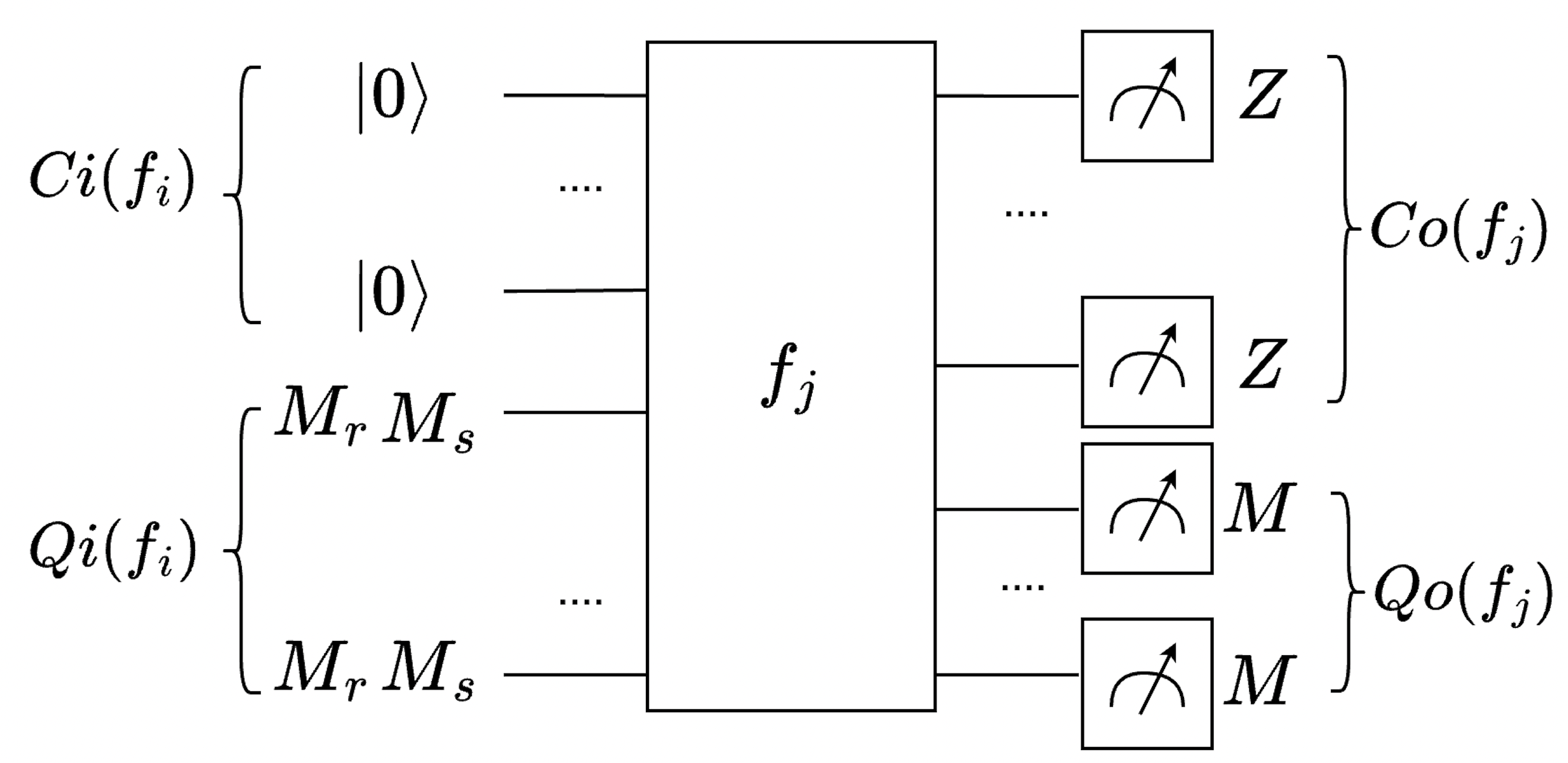} 
    \caption{An Example of Classical/Quantum Input/Output: The state initialized as \(\ket{0}\) represents the classical input $C_i(f_i)$, while states initialized as \(M_r\) or \(M_s\) are quantum inputs $Q_i(f_i)$. A qubit measured in the \(Z\) basis constitutes a classical output $C_i(f_i)$, whereas measurements in several bases, denoted by \(M\), are considered quantum outputs $Q_o(f_i)$.}
    \label{fig:QinQout} 
\end{figure}

Let us start from a simple state of a three-qubit quantum circuit, as depicted in Fig.~\ref{fig:circuit_cut_backgroud}. The state of the circuit can be expressed as \( \rho = VU\ket{000}\langle000|U^\dagger V^\dagger \). Here, \(V\) acts on the 1st and 2nd qubits, and \(U\) acts on the 2nd and 3rd qubits. For simplicity, although it should be \(U \otimes I\) and \(I \otimes V\), we use \(U\) and \(V\) to denote these operations respectively.
The cut, marked as a red cross mark in Fig.~\ref{fig:circuit_cut_backgroud}, separates the circuit into two 2-qubit fragments associated with the unitary operations $U$ and $V$ respectively. As detailed in \cite{peng2020simulating, tang2021cutqc}, a single qubit's state $\rho'$ can be decomposed as 
\begin{equation}
\rho' = \frac{1}{2}(\text{Tr}(I\rho')I + \text{Tr}(X\rho')X + \text{Tr}(Y\rho')Y + \text{Tr}(Z\rho')Z),
\end{equation}
Following this decomposition, the example circuit can be represented as:
\begin{equation}
\rho = \frac{1}{2} \sum_{M \in \mathcal{B}} \text{tr}_2 (M U\ket{00}\langle00|U^\dagger) \otimes V(M \otimes \ket{0}\langle0|)V^\dagger,
\label{eq:three qubit example circuit cutting}
\end{equation}
where $\mathcal{B}=\{I, X, Y, Z\}$ and $\text{tr}_i$ denotes the partial trace on the $i$-th qubit.
Denote the states for fragments $f_1$ and $f_2$ (parameterized by basis elements $M \in \mathcal B$) as
\begin{align}
    \rho_{f_1}(M) &= \text{tr}_2 (M U\ket{00}\langle00|U^\dagger), \\
    \rho_{f_2}(M) &= V(M \otimes \ket{0}\langle0|)V^\dagger,
\end{align}
the state of the circuit can be rewritten as the tensor product of states
\begin{equation}
\rho = \frac{1}{2} \sum_{M \in \mathcal{B}} \rho_{f_1}(M) \otimes \rho_{f_2}(M).
\end{equation}
Note that the partial trace operation in $\rho_{f_1}$ corresponds to a measurement, whereas the operator $M$ is used to indicate the initialization $M_r$ and $M_s$ for $\rho_{f_2}$. Moreover, we adopt the terminology from Ref.~\cite{Perlin_2020} and view circuit fragments as quantum channels. Therefore, we will refer to parts of fragments as classical/quantum input/output; see Fig.~\ref{fig:QinQout} for an illustration. 

For a more general case involving \( k \) cuts that produce \( N_f \) fragments \( \{f_1, f_2,...,f_{N_f}\} \), we define \( M(f_j) = Q_o(f_j) \otimes Q_i(f_j) \), where \( Q_o(f_j) \) and \( Q_i(f_j) \) represent the quantum output and input observables, respectively. We denote the classical output state corresponding to fragment \( f_j \) and operator $M(f_j)$ by \( \rho_{f_j}(M(f_j)) \), and the unitary operator within the circuit corresponding to  $j$-th fragment (ignoring initializations and measurements) as \( U_j \). The fragment \( f_j \) contains \( n_j \) qubits, which comprise \( n^{qi}_j \) quantum input qubits, \( n^{qo}_j \) quantum output qubits, \( n^{ci}_j \) classical input qubits, and \( n^{co}_j \) classical output qubits. \( Q_o(f_j) \) and \( Q_i(f_j) \) can be represented as a  \( n^{qo}_j \) and \( n^{qi}_j \) length Pauli string.
It is possible that \( n^{qo}_j \), \( n^{qi}_j \), \( n^{ci}_j \), and \( n^{co}_j \) could sometimes be zero; however, the relationship \( n_j = n^{qo}_j + n^{co}_j = n^{qi}_j + n^{ci}_j \) must be satisfied. 
For convenience, we denote the quantum input, quantum output, classical input, and classical output qubit index set for \( j \)-th fragment as \( \mathbf{N}^{qi}_j \), \( \mathbf{N}^{qo}_j \), \( \mathbf{N}^{ci}_j \), and \( \mathbf{N}^{co}_j \), respectively. For example, \( \mathbf{N}^{qi}_j \) should contain $n^{qi}_j$ qubit indices.  By \( \bigotimes^{\mathbf{N}^{ci}_j} \), we imply the tensor product of all qubits within the index set \( \mathbf{N}^{ci}_j \). Additionally, \( \text{tr}_{\mathbf{N}^{qo}_j} \) denotes the operation of taking the trace over all qubits indexed in \( \mathbf{N}^{qo}_j \).
The subcircuit \( \rho_{f_j}(M(f_j)) \) can be formally expressed as
\begin{equation}
\rho_{f_j}(M(f_j)) =\frac{1}{2^{n^q_j}} \text{tr}_{\mathbf{N}^{qo}_j} \left(Q_o(f_j)U_j Q_i(f_j) \otimes \ket{0}\langle0|^{\bigotimes \mathbf{N}_{ci}} U_j^\dagger\right)
\label{eq:subcircuit}
\end{equation}
where \(n^q_j = n^{qo}_j + n^{qi}_j\), the sum of the number of quantum input and output qubits.
The final uncut circuit can be represented as
\begin{equation}
\rho = \frac{1}{2^k} \sum_{M \in \mathcal{B}^k} \bigotimes_{j=1}^{N_f} \rho_{f_j}(M(f_j)).
\label{Eq circuit reconstrution}
\end{equation}
Note that the notation used in subsequent sections is derived from this section.

\subsection{MUBs-Based Grouping and Simultaneous Measurement}
\label{Simultaneous Measurement}
Simultaneous measurement is a technique that allows for the measurement of multiple observables in a circuit simultaneously, utilizing MUBs-based grouping. This is accomplished by organizing all observables into commuting groups, also known as commuting families in some literature. These groups comprise Pauli operators that commute with each other, allowing all observables within a group to be measured simultaneously in a single session. This approach can significantly reduce the number of measurements needed. It has been widely used in variational quantum eigensolver\cite{gokhale2019minimizing}, quantum phase estimation~\cite{cleve1998quantum}, and classical shadow tomography \cite{wang2023classical}. We also want to simplify circuit cutting via simultaneous measurement.

Let us start with the definition of the commutator. The commutator of two elements $g$ and $h$ is defined 
\begin{equation}
[g,h] = gh - hg.
\end{equation}
When $[g,h]=0$, it indicates that $g$ and $h$ commute. In quantum mechanics, when Hermitian observables commute, it implies that they share a common set of eigenstates \cite{shankar2012quantum}, \cite{horn2012matrix}. Thus, by measuring this set of eigenstates, we can simultaneously obtain information for all members in the commuting group. In information science, mutually unbiased bases (MUBs) \cite{wootters1986quantum, schwinger1960unitary, ivonovic1981geometrical, wootters1989optimal, lawrence2002mutually} guide the partitioning of $n$-qubit Pauli operators into optimal commuting groups, known as 'MUBs-based grouping'. The optimal partition of $\{ (I, X, Y, Z)^{\otimes n} \setminus {I^{\otimes n}} \}$, a set of $4^n-1$ Pauli operators, is $2^n+1$ groups, each containing $2^n-1$ Pauli operators.

For example, for 1-qubit Pauli strings, the commuting groups are trivial $\{X\},\{Y\},\{Z\}$, because $[X,X]=[Y,Y]=[Z,Z]=0$. In the two-qubit case, the optimal commuting groups are $\{XI,IX,XX\}$, $\{ZI,IZ,ZZ\}$, $\{YI,IY,YY\}$, $\{XI,IX,XX\}$, $\{YZ,ZX,XY\}$ and $\{XY,YZ,ZX\}$. For an abitrary $n$-qubit system, we utilize the method from \cite{gokhale2019minimizing} to find optimal commuting groups. It firstly constructs a graph where each node represents one of the $4^n-1$ Pauli strings. Nodes are connected if their corresponding Pauli operators commute. We then iteratively find $2^n+1$ cliques in this graph until all $4^n-1$ Pauli operators are covered. Lastly, $I^{\otimes n}$ can be assigned to any group.

\begin{figure}[htbp]
    \centering
    \includegraphics[width=0.6\linewidth]{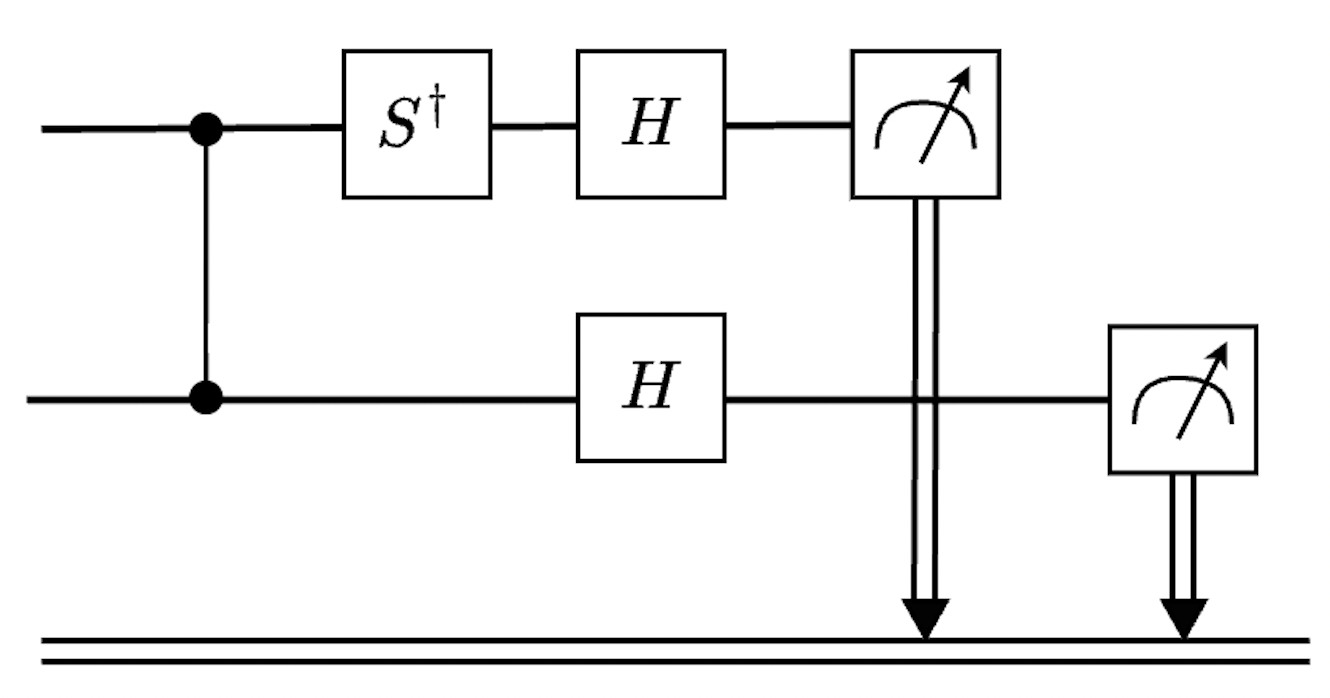} 
    \caption{Transformation circuit of group $\{XY,YZ,ZX\}$. After transforming the circuit, running it once yields results for \(\{XY, YZ, ZX\}\). Measuring the first qubit, the second qubit, and both the first and second qubits correspond to results for \(YZ\), \(ZX\), and \(-XY\), respectively.}
    \label{fig:figure5} 
\end{figure}

After partitioning, we need to perform simultaneous measurement for all observables within a group. Simultaneous measurement is straightforward if the shared eigenstates are tensor products of pure states on each qubit. For instance, in the group $\{XI, IX, XX\}$, each qubit can be easily measured in the $X$ basis. However, not all eigenstates can be decomposed into pure state, some groups have entangled eigenstates. For example in two qubits commuting group $\{YZ,ZX,XY\}$ has entangled eigenbasis $\{|\Phi^{+}\rangle, |\Phi^{-}\rangle, |\Phi^{+}\rangle, |\Psi^{-}\rangle\}$
$$|\Phi^{+}\rangle = \frac{1}{\sqrt{2}}\ket{0}\ket{+i}+\frac{1}{\sqrt{2}}\ket{1}\ket{-i}$$ 
$$|\Phi^{-}\rangle = \frac{1}{\sqrt{2}}\ket{0}\ket{+i}-\frac{1}{\sqrt{2}}\ket{1}\ket{-i}$$
$$|\Psi^{+}\rangle = \frac{1}{\sqrt{2}}\ket{0}\ket{-i}+\frac{1}{\sqrt{2}}\ket{1}\ket{+i}$$ 
$$|\Psi^{-}\rangle = \frac{1}{\sqrt{2}}\ket{0}\ket{-i}-\frac{1}{\sqrt{2}}\ket{1}\ket{+i}$$
It is similar for group $\{XY,YZ,ZX\}$. Thus, we need to transform the eigenbasis into $Z$ basis so we can get the output by directly measure them \cite{gottesman1997stabilizer}. The way to measure on basis is applying a transformation circuit before the place we are measuring. The transformation circuit can be build in different way \cite{seyfarth2011construction},\cite{reggio2023fast},\cite{aaronson2004improved}.
As an example, if we want to build a transformation circuit for group $\{XY,YZ,ZX\}$, we can first apply a $CZ$ gate between two qubits,  $H$, $S^\dagger$ on the first qubit and $H$ on the second qubit, shown in Fig.~\ref{fig:figure5}. To verify this, we can use a basis transformation \( TXYT^\dagger \), where \( T \) is the transformation circuit.
In this example, $T= (H \otimes I) \cdot (S^\dagger \otimes H) \cdot CZ$ and we have
$$(H \otimes I) \cdot (S^\dagger \otimes H) \cdot CZ  \cdot YZ \cdot [(H \otimes I) \cdot (S^\dagger \otimes H) \cdot CZ]^\dagger = ZI$$
$$(H \otimes I) \cdot (S^\dagger \otimes H) \cdot CZ \cdot ZX \cdot [(H \otimes I) \cdot (S^\dagger \otimes H) \cdot CZ]^\dagger = IZ$$
$$(H \otimes I) \cdot (S^\dagger \otimes H) \cdot CZ \cdot XY \cdot  [(H \otimes I) \cdot (S^\dagger \otimes H) \cdot CZ]^\dagger =-ZZ$$ 
Thus, measuring on the first qubit, the second qubit, and both the first and second qubit correspond to results for $YZ$,$ZX$, and $-XY$. The worst case of transformation circuit requires $O(n^2)$ gates \cite{seyfarth2011construction}.

\section{Converting Quantum input to Quantum output}
\label{Ancilla assist}
For quantum input, we prepare eigenstates of each Pauli basis at the beginning of a circuit. Initializing states for all bases requires a total of $6^{n^{qi}}$ states for $n^{qi}$ quantum input qubits. Taking one cut as an example, the states to be prepared include $\ket{0}$, $\ket{1}$, $\ket{+}$, $\ket{-}$, $\ket{+i}$, and $\ket{-i}$. The initialization requirement can be reduced to preparing only $\ket{0}$, $\ket{1}$, $\ket{+}$, and $\ket{+i}$ states by substituting the identity operation with $\ket{0}$ and $\ket{1}$, though $4^{n^{qi}}$ initial states are still required \cite{tang2021cutqc}.

Drawing inspiration from quantum process tomography \cite{levy2024classical,altepeter2003ancilla}, applying a random Pauli string at the front of the circuit is analogous to entangling each qubit with an ancilla qubit in a Bell state and then applying a measurement on that Pauli basis. We provide a formula and example Fig.~\ref{fig:figure1} below.

\begin{figure}[htbp]
    \centering
    \includegraphics[width=\linewidth]{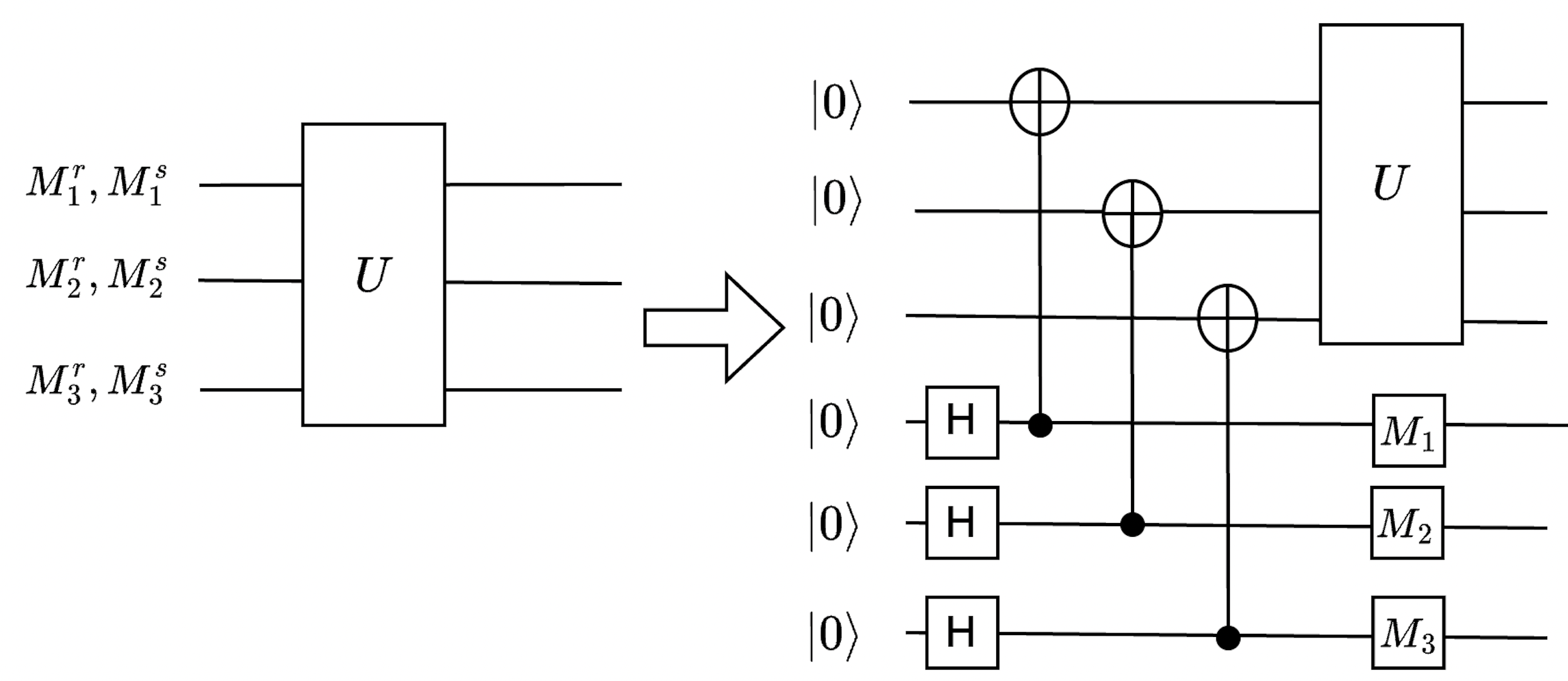} 
    \caption{An Example of Ancilla-Assisted Initialization. The left circuit involves initializing each qubit separately with \( M_i^r \) and \( M_i^s \), while the right circuit involves measuring each qubit in the \( M_i \) basis. And \( M_i = rM_i^r + sM_i^s \). The circuit on the left needs to be executed eight times to get the expectation value. However, by running the right circuit once, we can obtain the expectation value, which is much more convenient.} 
    \label{fig:figure1} 
\end{figure}

Assuming we have an \( n \)-qubit circuit \( U \). All \( n \) qubits are quantum input qubits, which means \( n = n^{qi} \) (\( n = n^{qi} \) is used only in this section). We need to initialize different eigenstate $\ket{u}_i$ and $\ket{v}_i$ on $i$-th qubit, where $\ket{u}_i$ and $\ket{v}_i$ are eigenstates of $M_i$, and $M=\bigotimes_{i=1}^{n}M_i= \bigotimes_{i=1}^{n}(r\ket{u} \langle u|_i + s\ket{v} \langle u|_i) = \bigotimes_{i=1}^{n}(rM_i^r + sM_i^s) $. 
Just like ancilla-assisted quantum process tomography, we can rewrite $\frac{1}{2^n} \text{tr}\left(O \cdot [ U^\dagger \bigotimes_{i=1}^{n}M_i U ] \right)$ via entangling each qubit with an ancilla-qubit in a bell state and then measure ancilla-qubit on $M$ basis, as shown below:
\begin{align}
& \frac{1}{2^n} \text{tr}\left(O \cdot [ U\bigotimes_{i=1}^{n}M_i U^\dagger ] \right) \notag \\
&= \frac{1}{2^n} \text{tr}\left(O \cdot [ U \bigotimes_{i=1}^{n} (r |u\rangle \langle u|_i + s  |v\rangle \langle v|_i ) U^\dagger ] \right) \notag \\
&= \frac{1}{2^n} \text{tr}\left( U^\dagger  O  U \cdot  \bigotimes_{i=1}^{n} (r |u\rangle \langle u|_i + s  |v\rangle \langle v|_i )  \right) \notag \\
&= \text{tr}\left(U^\dagger  O  U \otimes M \cdot  \bigotimes_{i=1}^{n} \left(\frac{1}{2}|uu\rangle \langle uu|_{i,i'} + \frac{1}{2}|vv\rangle \langle vv|_{i,i'}\right) \right) \notag \\
&= \text{tr}\left(O \otimes M \cdot  \hat{U} \bigotimes_{i=1}^{n}  \left(\frac{1}{2}|uu\rangle \langle uu|_{i,i'} + \frac{1}{2}|vv\rangle \langle vv|_{i,i'} \right) \hat{U^\dagger }  \right) \label{Eq: Ancilla qubit assist}
\end{align}
where $\hat{U} = U \otimes I$ and $\frac{1}{2}(|uu\rangle \langle uu|_{i,i'} + |vv\rangle \langle vv|_{i,i'})$ is bell state on $i$-th qubit and $i'$-th qubit. Following the qubit ordering, $i'$ may be $i+n$. This strategy simplifies the implementation as the one circuit can be used for all initialization schemes. 

\section{Efficient Circuit Wire Cutting based on commuting groups}
\label{se:Circuit Wire Cutting based on commuting grouping}
Currently, CutQC~\cite{tang2021cutqc} represents the most promising work in circuit cutting. It encompasses the most widely used decomposition method for circuit cutting. Thus, in this paper, we compare our circuit cutting technique with CutQC.

In CutQC, each density matrix at a cutting location need to be decomposed into four Pauli operators. When multiple cuts are introduced, the number of decomposed Pauli operators scales exponentially. Drawing inspiration from \cite{gokhale2019minimizing}, we can apply the MUBs-based grouping on $4^{n^q_j}$ Pauli operators as discussed in Section~\ref{Simultaneous Measurement}. This approach allows for partitioning the set \(\{ (I, X, Y, Z)^{\otimes n^q_j} \setminus I^{\otimes n^q_j}\} \), consisting of \( 4^{n^q_j}-1 \) Pauli operators, into \( 2^{n^q_j}+1 \) commuting groups, each group contains \( 2^{n^q_j}-1 \) Pauli operators. By utilizing this method, information can be obtained more efficiently.\\
We provide a more detailed derivation as follows. More specifically, recall from Eq.~\ref{eq:subcircuit} that the subcircuit of $j$-th fragment with operator $M(f_j)$ can be expressed as:
\begin{align}
      \rho_{f_j}(M(f_j)) =\frac{1}{2^{n^q_j}} \text{tr}_{\mathbf{N}^{qo}_j} \left(Q_o(f_i)U Q_i(f_i) \otimes \ket{0}\langle0|^{\bigotimes \mathbf{N}^{ci}_j} U^\dagger \right)
\end{align}
Then, we convert quantum input to quantum output by adding \(n^{qi}_j\) ancilla qubits as discussed in Sec.~\ref{Ancilla assist}. This yields
\begin{align}
     \rho_{f_j}(M(f_j)) = & \frac{1}{2^{n^{qo}_j}} \text{tr}_{\mathbf{N}^{qo}_j} \left( Q_o(f_j) \otimes Q_i(f_j) \cdot \ket{\phi} \langle \phi| \right) \notag\\
\end{align}
where the state \(\ket{\phi}\) is given by
\[
\ket{\phi} = \hat{U} \left( |0\rangle^{\otimes \mathbf{N}^{ci}_j} \otimes \frac{1}{\sqrt{2^{n^{qi}_j}}} (|00\rangle_{i, i+n_j} + |11\rangle_{i, i+n_j})^{\otimes \mathbf{N}^{qi}_j} \right)
\]
, \(\hat{U} = U \otimes I^{\otimes \mathbf{N}^{qi}_j}\) and \(\frac{1}{\sqrt{2^{n^{qi}_j}}} (|00\rangle_{i, i+n_j} + |11\rangle_{i, i+n_j})^{\otimes \mathbf{N}^{qi}_j}\) is the Bell state entangling the \(i\)-th qubit and the \((i+n_j)\)-th qubit. For convenient we denote the quantum input qubit index set after conversion as $\bar{\mathbf{N}}^{qi}_j$.

\begin{figure}[htbp]
    \centering
    \begin{flushright}
    \includegraphics[width=\linewidth]{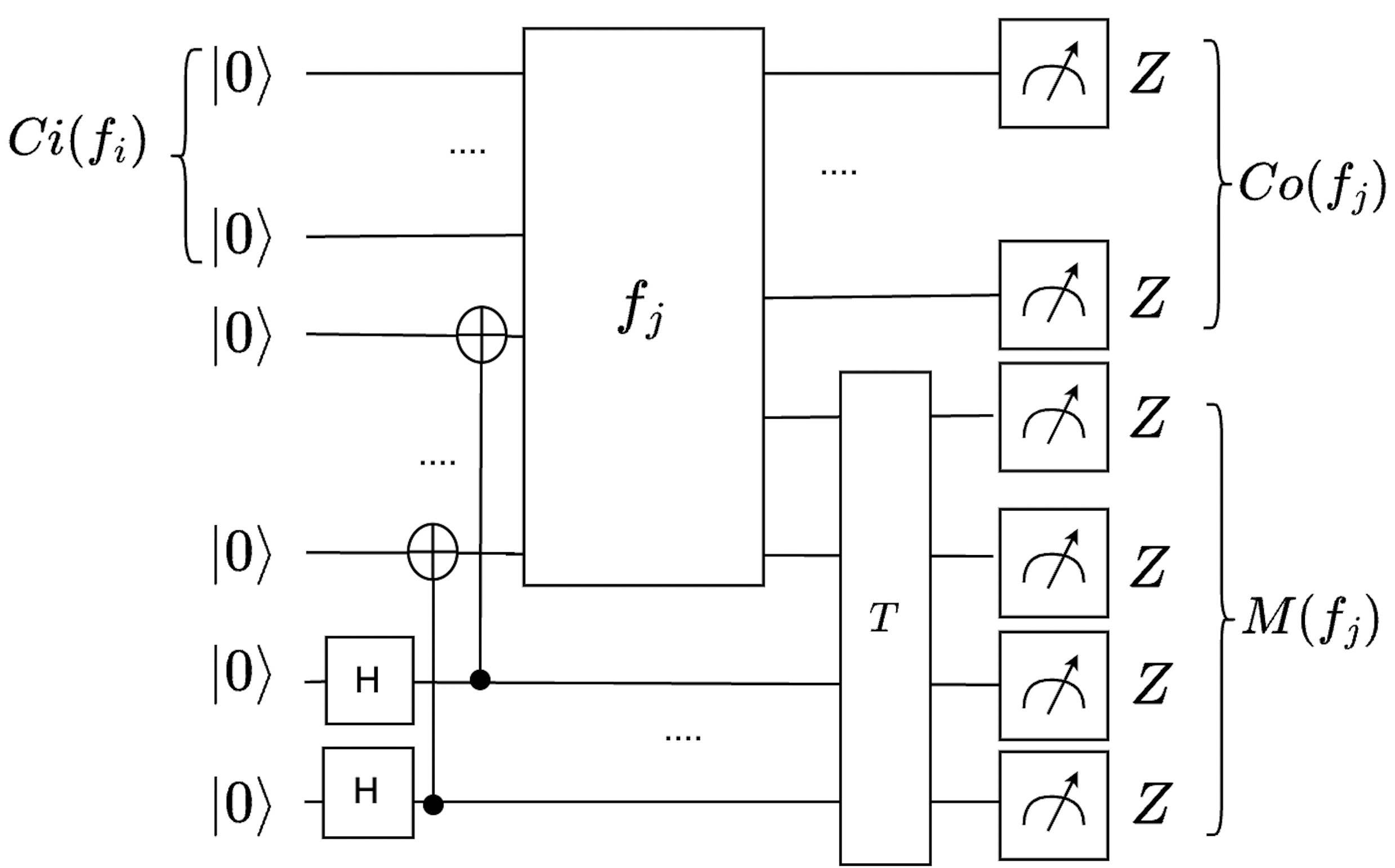} 
    \caption{Example of a subcircuit converted from Fig.~\ref{fig:QinQout}. With the use of an ancilla qubit, we can convert quantum inputs into quantum outputs. This allows us to perform MUBs-based grouping on  $M(f_j) = Qo(f_j) \otimes Qi(f_j)$.}
     \label{fig:figure4} 
    \end{flushright}
\end{figure}

We can then apply MUBs-based grouping on $4^{n^q_j}$ Pauli operators and perform simultaneous measurement on \(\bar{\mathbf{N}}^{qi}_j \cup \mathbf{N}^{qo}_j\) qubits within each group. Denote $G(f_j)$ as one of $2^{n^q_j}+1$ commuting groups. Denote $T$ as a transformation circuit (Sec.~\ref{Simultaneous Measurement}) corresponding $G(f_j)$. We have
\begin{align}
    \rho_{f_j}(G(f_j)) = & \frac{1}{2^{n^{qo}_j}} \text{tr}_{\mathbf{N}^{qo}_j}\left( Z^{\otimes \bar{\mathbf{N}}^{qi}_j \cup \mathbf{N}^{qo}_j} \cdot \hat{T} \ket{\phi} \langle \phi| \hat{T}^{\dagger} \right) \notag \\
\end{align}
where \(\hat{T} = I^{\otimes \mathbf{N}^{co}_j} \otimes T\). Now, we can measure on the basis \( Z^{\otimes (\bar{\mathbf{N}}^{qi}_j \cup \mathbf{N}^{qo}_j)} \) to get results for the commuting group. Any subcircuit state \( \rho_{f_j}(M(f_j)) \) can be obtained through post-processing of \( \rho_{f_j}(G(f_j)) \), if \( M(f_j) \in G(f_j) \). A sample example is shown in Fig.~\ref{fig:figure4}. 

\begin{figure*}
    
  \centering
  \begin{subfigure}{0.33\textwidth}  
    \includegraphics[width=\textwidth]{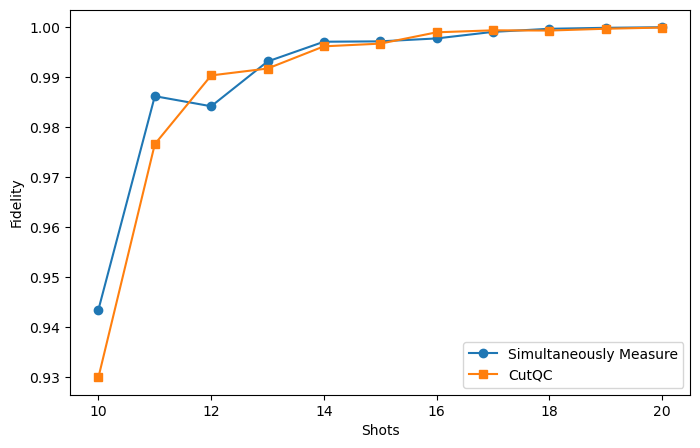}
    \caption{3Fragments, 2Cuts, 6 qubits}
    \label{fig:cut2=2}
  \end{subfigure}%
  \begin{subfigure}{0.32\textwidth}  
    \includegraphics[width=\textwidth]{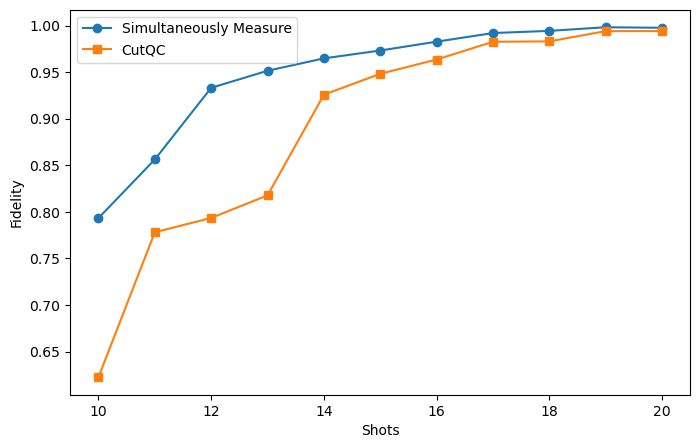}
    \caption{3Fragments, 3Cuts, 7 qubits}
    \label{fig:cut2=3}
  \end{subfigure}%
  \begin{subfigure}{0.32\textwidth}  
    \includegraphics[width=\textwidth]{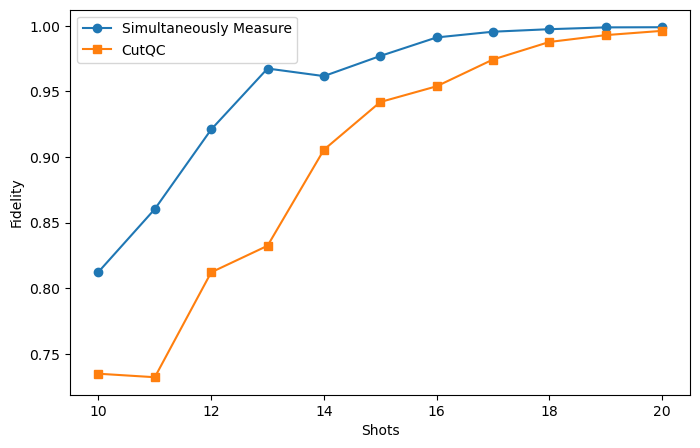}
    \caption{3Fragments, 4cuts, 8 qubits}
    \label{fig:cut2=4}
  \end{subfigure}%

  \caption{Circuit Cutting Reconstruction Results under Different Shots. In this experiment, we maintain the same total number of shots for each fragment. The x-axis represents the logarithmic number of shots, shots range from \(2^{10}\) to \(2^{20}\). The y-axis measures the fidelity of the reconstructed results. The orange line represents the reconstructed results from CutQC, while the blue line represents our efficient circuit cut method. It is evident that when more cuts are present in the circuit, our method achieves higher fidelity, provided the same shot budget.}
  \label{fig:test}
\end{figure*}
After simultaneously measuring on all $2^{n^q_j}+1$ commuting groups for each fragment $f_j$ and completing all $N_f$ fragments, we reconstruct the result of the uncut circuit using the Eq.\ref{Eq circuit reconstrution}.

This method maximally leverage the advantage of MUBs-based grouping. There is a straightforward benefit that the number of circuits to run decreases from $4^{n^{qi}_j}3^{n^{qo}_j}$ to $2^{n^{q}_j}+1 = 2^{n^{qi}_j+n^{qo}_j}+1$. However, we can see this method is not optimal for all kinds of fragments considering the shot budget. We will provide more analysis in the next section.



    

\section{Shot Analysis and Trade-offs}
\label{se:pro con}
\subsection{Shot Analysis}
Let $S(n)$ be the number of shots required to estimate an observable from an $n$-qubit circuit. According to CutQC, the total number of shots required for the $j$-th fragment should be
\begin{equation}
    4^{n^{qi}_j}3^{n^{qo}_j}S(n^{qo}_j+ n^{co}_j). \label{Eq. CutQC shots}
\end{equation}
The exponential factors is the number of subcircuits (initialization-measurement pair) to run. We first replace quantum inputs with quantum outputs, which changes the total number of shots to
\begin{align}
    3^{n^{qi}_j + n^{qo}_j}S(n^{qi}_j + n^{qo}_j + n^{co}_j).
\end{align}
Then, exploiting MUBs-based grouping further reduces the total number of circuits to execute polynomially, yielding the sampling complexity: 
\begin{align}
    (2^{n^{qi}_j+n^{qo}_j}+1)S(n^{qi}_j + n^{qo}_j+ n^{co}_j).
\end{align}

We can perform a more detailed analysis with explicit form of $S$. Typically, it is reasonable to set that 
\begin{align}
S(n^{qi}_j + n^{qo}_j+ n^{co}_j) = 2^{n^{qi}_j}S(n^{qo}_j+ n^{co}_j)
\end{align}
as there are $2^{n^{qi}_j}$ possible states from at the quantum output qubit need to be fully measure. Thus, we can reform our method results as:
\begin{align}
 (4^{n^{qi}_j} 2^{n^{qo}_j}+2^{n^{qi}_j})S(n^{qo}_j+ n^{co}_j)
\end{align}
Compared to Eq.~\ref{Eq. CutQC shots}, our method polynomially better than CutQC when \(n^{qo}_j > 0\) and performs slightly worse when there are no quantum outputs. Thus, we give an Algorithm \ref{algorithm} that applies the CutQC method when there are no quantum outputs, and utilizes our efficient circuit cutting technique when there is quantum outputs. The algorithm can guarantee that all fragments achieve shot savings.

\begin{algorithm}
\label{algorithm}
\caption{Shot Saving Efficient Circuit Cutting}
\begin{algorithmic}[1] 
\State \textbf{Input:} fragments set $\mathcal{F} = \{f_j \text{ for all } j\}$, observable $O$
\State \textbf{Output:} Expectation $\text{tr}(O\rho)$
\For{$f_i \in \mathcal{F}$}
    \State Get all information about $Q_o(f_i), Q_i(f_i)$, $n^{qo}_j$, $n^{qi}_j$
    \If{$n^{qo}_j > 0$}
        \State Apply MUBs-based grouping, run $2^{n^{q}_j+n^{qi}_j}+1$ subcircuits
        \State Post-processing based on MUBs-based grouping.
    \Else
        \State 
        Apply CutQC initialization, run  $4^{n^{qi}_j}$ subcircuits.
        \State Post-processing based on CutQC.
    \EndIf

\EndFor
\State Reconstruct the result of the uncut circuit
\end{algorithmic}
\label{algorithm}
\end{algorithm}
\subsection{Trade-offs Regarding Circuit Width and Depth}
As discussed in Sec.~\ref{Simultaneous Measurement}, to perform simultaneous measurement, we need to apply a transformation circuit to the measurement qubits. The maximum size of this transformation circuit is \( O({n^q_j}^2) \), where \( {n^q_j}^2 = (n^{qo}_j + n^{qi}_j)^2 \) represents the squaure of the total number of quantum input and output qubits. For smaller fragments, this process may not be well-suited to current noisy devices, as it introduces ancilla qubits, additional CNOT gates, and a transformation circuit. The number of gates in a fragment should be \( \Omega({n^q_j}^2) \) to ensure that the impact of the transformation circuit is negligible. Additionally, introducing $n^{qi}_j$ more ancilla qubits lead to an increase in circuit width.

\section{Experiment}
\label{se:experiment}

\begin{figure}[htbp]
    \centering
    \includegraphics[width=0.8\linewidth]{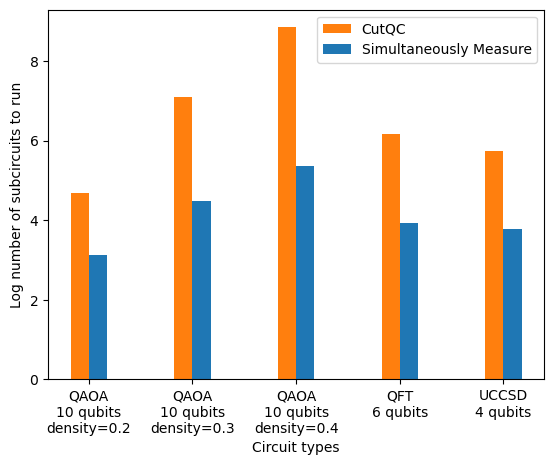} 
    \caption{Number of Subcircuits in Different Types of Circuits. We test the division of QAOA, QFT, and UCCSD circuits into halves by cutting n-qubit circuits. The cutting locations are determined using the MIP model from \cite{tang2021cutqc}. The x-axis categorizes the circuit types, and the y-axis represents the logarithmic number of subcircuits. The orange bars represent the number of subcircuits from CutQC, while the blue bars are from our efficient circuit cutting method. It is evident that our method significantly reduces the polynomial number of subcircuits required.}
    \label{fig:number subcircuit} 
\end{figure}

We first demonstrate that our method significantly reduces the number of subcircuits required for execution. As shown in Fig.~\ref{fig:number subcircuit}, we observe the number of subcircuits necessary for various circuits. We utilize the Mixed-Integer Programming (MIP) 
model from CutQC \cite{tang2021cutqc} to determine cutting locations. The maximum subcircuit width is set to \(n/2 + 1\), where $n$ is number of qubit in the uncut circuit. We explore fragment numbers ranging from 1 to 10, selecting the configuration that yields the minimum number of subcircuits. The circuits under investigation include quantum approximate optimization algorithms (QAOA) with randomly generated graphs of different densities, as well as quantum Fourier transform (QFT) and Unitary Coupled-Cluster Single and Double excitations variational form (UCCSD )\cite{barkoutsos2018quantum}.

To assess the overhead in terms of shots, we employ randomly generated circuits as benchmarks. We sampled several circuits randomly as fragments and combined them to create our test uncut circuits. The tests were carried out with 2, 3, and 4 cuts. All circuits were executed on the Aer Simulator without introducing any hardware noise.
As shown in Fig.~\ref{fig:cut2=2}, the performance with 2 cuts does not differ significantly from CutQC. However, as depicted in Fig.~\ref{fig:cut2=3} and \ref{fig:cut2=4}, our method converges faster than CutQC as shots increasing, indicating a potential reduction in the number of shots required.




\section{Conclusion}
\label{conclusion}
In this paper, we introduce a new method for executing circuit cutting that reduces the shot overhead associated with quantum measurements. Our method first converts all quantum inputs into outputs using ancilla-assisted initialization. We then use a transformation circuit composed the quantum inputs and outputs to enable simultaneous measurement within a commuting group. Our analysis focuses on determining under which conditions MUBs-based grouping yields improvements, and we give an algorithm to guarantee shot savings. We conclude by demonstrating that our method outperforms previous techniques on saving shots and reducing number of quantum circuits to run.

\bibliographystyle{IEEEtran}
\bibliography{Exp1,cite}

\end{document}